\documentclass[runningheads]{llncs}
\usepackage{float}
\usepackage{graphicx}
\usepackage{cite}
\usepackage{amssymb}
\usepackage{multirow}
\usepackage{tikz}
\usepackage{booktabs}
\usepackage{amsmath}
\usepackage{caption}
\usepackage[colorlinks,
            linkcolor=blue,
            anchorcolor=blue,
            citecolor=blue]{hyperref}
\usepackage{hyperref}
\usepackage{marvosym}

\makeatletter
\newcommand\figcaption{\def\@captype{figure}\caption}
\newcommand\tabcaption{\def\@captype{table}\caption}
\makeatother
 
\begin{document}
\title{H2ASeg: Hierarchical Adaptive Interaction and Weighting Network for Tumor Segmentation in PET/CT Images}
\titlerunning{Hierarchical Adaptive Interaction and Weighting Network}

%
%
\author{Jinpeng Lu\inst{1}, Jingyun Chen\inst{1}, Linghan Cai\inst{2}\textsuperscript{(\Letter)}, Songhan Jiang\inst{2}, \\ \and Yongbing Zhang\inst{2}\textsuperscript{(\Letter)}}
%
\authorrunning{J. Lu and J. Chen et al.}
%
\institute{{School of Science, Harbin Institute of Technology (Shenzhen), \\ Shenzhen 518055, China \and School of Computer Science and Technology, Harbin Institute of Technology
(Shenzhen), Shenzhen 518055, China}\\
\email{\href{mailto: cailh@buaa.edu.cn}{cailh@buaa.edu.cn}, \href{mailto: ybzhang08@hit.edu.cn}{ybzhang08@hit.edu.cn}}
}

\maketitle              
\renewcommand{\thefootnote}{}
\footnotetext{J. Lu and J. Chen—Contributed equally to this work.}
\begin{abstract}
Positron emission tomography (PET) combined with computed tomography (CT) imaging is routinely used in cancer diagnosis and prognosis by providing complementary information. Automatically segmenting tumors in PET/CT images can significantly improve examination efficiency. Traditional multi-modal segmentation solutions mainly rely on concatenation operations for modality fusion, which fail to effectively model the non-linear dependencies between PET and CT modalities. Recent studies have investigated various approaches to optimize the fusion of modality-specific features for enhancing joint representations. However, modality-specific encoders used in these methods operate independently, inadequately leveraging the synergistic relationships inherent in PET and CT modalities, for example, the complementarity between semantics and structure. To address these issues, we propose a Hierarchical Adaptive Interaction and Weighting Network termed H2ASeg to explore the intrinsic cross-modal correlations and transfer potential complementary information. Specifically, we design a Modality-Cooperative Spatial Attention (MCSA) module that performs intra- and inter-modal interactions globally and locally. Additionally, a Target-Aware Modality Weighting (TAMW) module is developed to highlight tumor-related features within multi-modal features, thereby refining tumor segmentation. By embedding these modules across different layers, H2ASeg can hierarchically model cross-modal correlations, enabling a nuanced understanding of both semantic and structural tumor features. Extensive experiments demonstrate the superiority of H2ASeg, outperforming state-of-the-art methods on AutoPet-II and Hecktor2022 benchmarks. The code is released at \href{https://github.com/JinPLu/H2ASeg}{https://github.com/JinPLu/H2ASeg}.

\end{abstract}

\section{Introduction}
\label{sec:introduction}
Positron emission tomography (PET) combined with computed tomography (CT) is considered the imaging modality of choice for the diagnosis, staging, and monitoring of treatment responses in various cancers \cite{bussink2011pet, mu2020non, ell2006contribution}. Benefiting from the sensitivity to high metabolic areas, PET has demonstrated excellent capabilities in localizing tumors. Nevertheless, the low resolution of PET images prevents radiologists from relying solely on PET for outlining tumors. Conversely, despite CT cannot detect metabolic activity, it provides detailed structural information. Hence, the integration of PET and CT yields rich complementary information that enables accurate presentation of tumor regions \cite{kapoor2004introduction}. In clinical practice, manual annotation is a time-consuming process, limiting the effect of PET/CT examinations. Fortunately, the development of deep learning \cite{wang2021transbts, xiang2022modality, andrearczyk2023automatic, zhang2022mmformer, nnFormer} presents new avenues for automatic tumor segmentation in PET/CT.

In the multi-modal segmentation task, early convolutional neural networks (CNNs) such as UNet-3D \cite{UNet-3D}, VNet \cite{VNet}, and ResUNet-3D \cite{ResUNet-3D} face limitations in capturing contextual information, resulting in high-metabolism organs, which have a similar presentation to tumor regions in PET scans, often being misidentified. Subsequent works (e.g., UNETR \cite{UNETR} and SwinUNETR \cite{SwinUNETR}) leverage Transformer \cite{ViT,shamshad2023transformers} to explicitly model long-range spatial dependencies, significantly improving tumor segmentation performance. However, these methods rely on a simplistic concatenation operation for modality fusion that fails to model the non-linear relationship between PET and CT modalities. Thus, they cannot fully exploit the complementarity of each imaging modality in tumor segmentation. 

Considering the specificity of different modalities, A2FSeg \cite{A2FSeg} tailors an encoder for each modality to derive modality-specific feature representations, and then adopts an adaptive fusion module to generate a joint feature representation for tumor segmentation. NestedFormer \cite{NestedFormer} investigates a modality-aware feature aggregation module to fuse high-level features and uses the fused features to adaptively select shallow features for better decoding. Despite significant advancements, these solutions \cite{A2FSeg, NestedFormer, li2020deep, zhao2018tumor} struggle to accurately segment tumors in PET/CT, as the modality-specific encoders are generally independent of each other, posing challenges in exploring the synergistic relations inherent in PET and CT modalities. To make use of the relation, Xue \textit{et al.} \cite{SDB} propose a multi-modal co-learning network that replaces the down-sampling blocks in the dual encoder with the shared down-sampling blocks to promote the feature interaction between PET and CT images. Although the intent behind this strategy makes sense, significant differences between PET and CT modalities often weaken the effectiveness of this module and make information transfer difficult.

To address these issues, we present a Hierarchical Adaptive Interaction and Weighting Network named H2ASeg for precise PET/CT tumor segmentation. Our motivation stems from the fact that during tumor annotation, radiologists repeatedly observe PET/CT images to roughly locate a tumor region, and subsequently use detailed information around the tumor to obtain the silhouette mask. We, therefore, argue that the hierarchical interaction of PET and CT images in a deep learning network is crucial for tumor localization, in addition to highlighting features related to the tumor for accurate boundary discrimination. To these ends, H2ASeg uses Modality-Cooperative Spatial Attention (MCSA) modules to execute modality interaction at different levels, in which the inter- and intra-window Bi-Directional Spatial Attention (BDSA) mechanisms bridge the connections between modalities globally and locally, enriching the complementary information. Meanwhile, a Target-Aware Modality Weighting (TAMW) module is developed to focus on tumor-related features for optimizing tumor boundaries. In a nutshell, our contributions can be summarized as follows. (1) We present an H2ASeg that hierarchically models the correlations between PET and CT, exploiting the complementary information for precise tumor segmentation. (2) We propose an MCSA module that can transfer valuable information across modalities at both local and global scales. Meantime, a TAMW module is developed to highlight tumor-related features from multi-modal features for segmentation refinement. (3) Extensive experiments demonstrate the superiority of H2ASeg, achieving state-of-the-art performance on two datasets.

\begin{figure}[t]
\includegraphics[width=\textwidth]{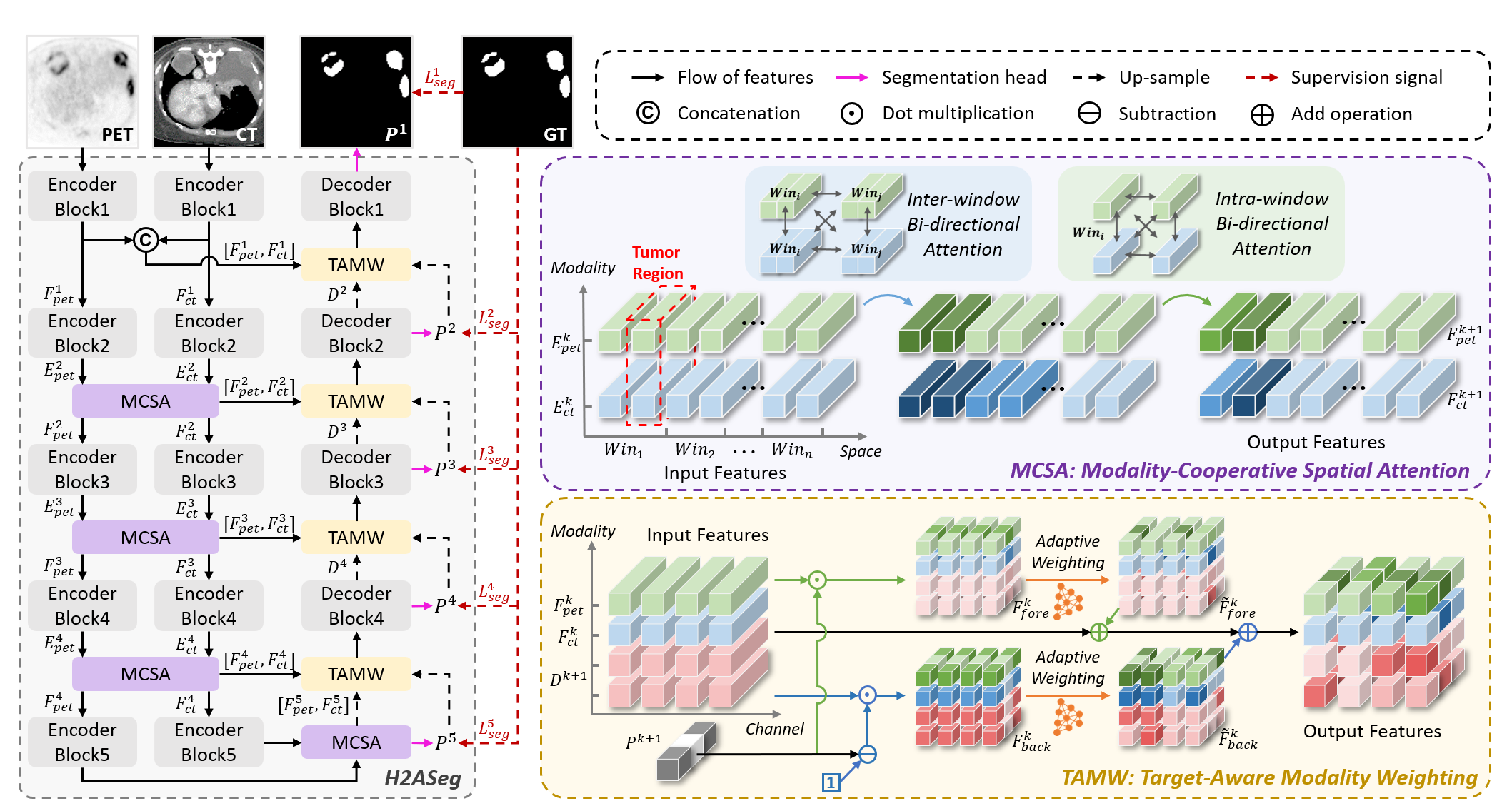}
\caption{Overview of the proposed H2ASeg, which is an encoder-decoder framework. MCSA modules perform modality interaction in the encoder. TAMW modules adjust multi-modal features for refining segmentation.}
\label{fig: Overview}
\end{figure}

\section{Methodology}
Fig. \ref{fig: Overview} illustrates the overall framework of \textbf{H2ASeg}, which adopts \textbf{two} modality-\textbf{A}daptive components into the \textbf{H}ierarchies of a dual-branch ResUNet-3D for tumor \textbf{Seg}mentation. At each level, MCSA executes feature interaction within and between modalities, collecting rich complementary information to improve feature representation. TAMW encourages the network to collect tumor-related features and then selects these features to refine tumor segmentation results.

\subsection{Modality-Cooperative Spatial Attention (MCSA)}
\begin{figure}[t]
\includegraphics[width=\textwidth]{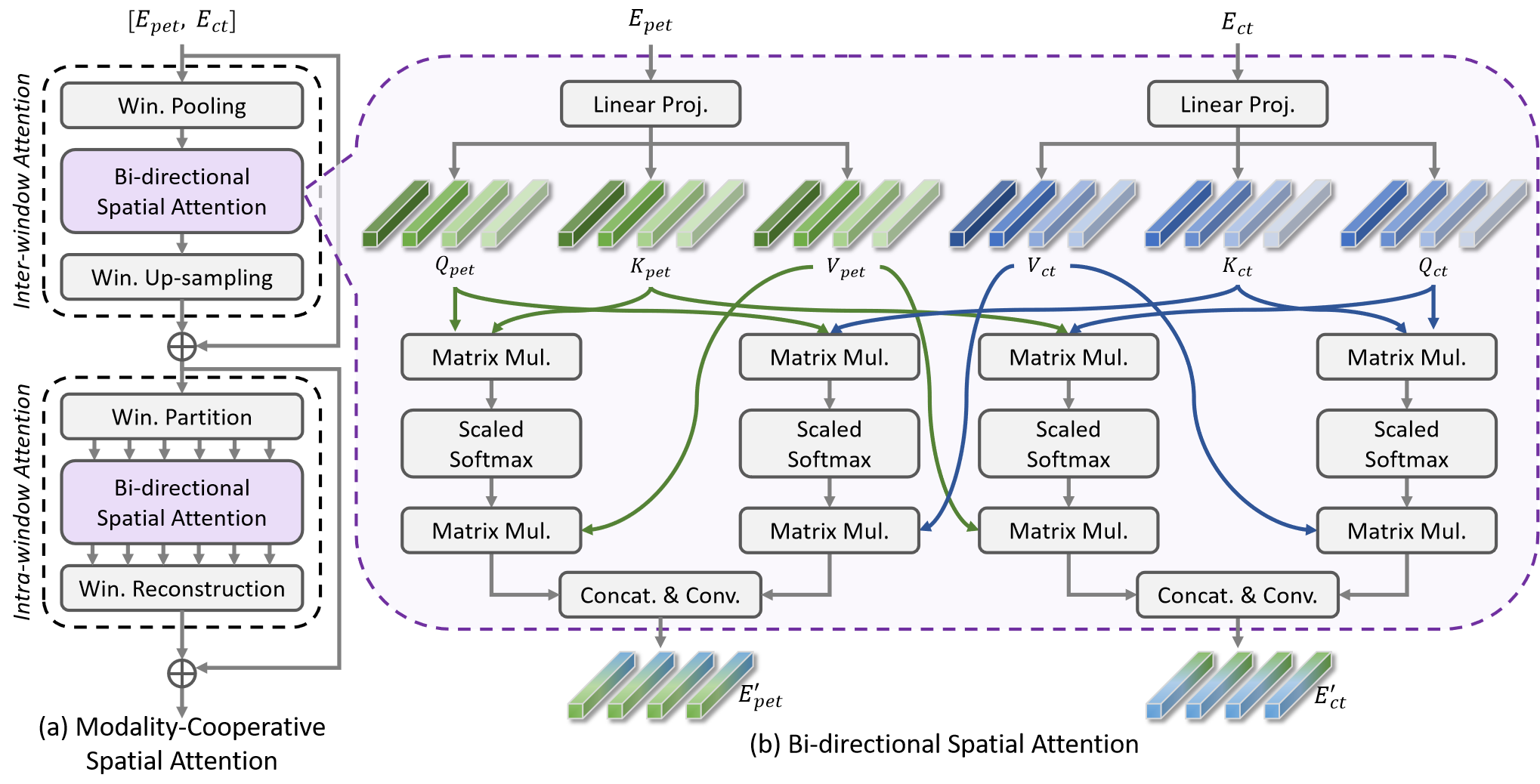}
\caption{Structure of modality-cooperative spatial attention module. (a) presents the overall architecture. (b) shows a bi-directional spatial attention mechanism. ``Win.'' is the abbreviation of window, ``Proj.'' is projection, ``Mul.'' means multiply, ``Concat. \& Conv." denotes concatenation and convolution.}
\label{fig: MCSA}
\end{figure}

Given the output features $E_{ct}^k$, $E_{pet}^k$ from the $k^{th}$ encoder block, MCSA treats them as two sets of 3D windows with the channel number of $C^{k}$ and the window size of $(H_{win}^k \times W_{win}^k \times D_{win}^k)$. As shown in Fig. \ref{fig: MCSA} (a), MCSA firstly executes inter-window attention to establish the long-range dependencies between windows and then captures the details through intra-window attention. During the process, a Bi-Directional Spatial Attention (BDSA) mechanism is used for feature interaction within and between modalities to collect complementary information.

Concretely, in inter-window attention, each window of the inputs is pooled into a token by a convolutional layer with the kernel size and stride equal to the window size $(H_{win}^k, W_{win}^k, D_{win}^k)$. In the process, the size of $E_{pet}^k$ and $E_{ct}^k$ changes from $(C^k\times H^k\times W^k\times D^k)$ to $(C^k\times \frac{H^k}{H_{win}^k}\times \frac{W^k}{W_{win}^k}\times \frac{D^k}{D_{win}^k})$. Subsequently, BDSA is used to model long-range dependencies across windows. After that, window up-sampling, a trilinear interpolation operation, reverses the above shape change to generate intermediate outputs $\hat{E}_{pet}^{k}$, $\hat{E}_{ct}^{k}\in \mathbb{R}^{C^k\times H^k\times W^k\times D^k}$.

In intra-window attention, window partition splits the feature into two ordered sets, $S^{k}_{pet}$, $S^{k}_{ct}\in \mathbb{R}^{N_h^k N_w^k N_d^k\times C^k\times H_{win}^k\times W_{win}^k\times D_{win}^k}$, of windows with preset shapes, where $N_h^k$, $N_w^k$, $N_d^k$ denote the number of windows on the height, width, and depth axes respectively. Then, BDSA is implemented in each window to enhance local details. Finally, the windows are rearranged in order through window reconstruction to obtain the final output $F_{pet}^{k}, F_{ct}^{k}\in \mathbb{R}^{C^k\times H^k\times W^k\times D^k}$.

\subsubsection{Bi-Directional Spatial Attention (BDSA).} 
Fig. \ref{fig: MCSA} (b) shows the structure of the BDSA mechanism, which can be simplified as two parallel unidirectional processes, using PET and CT as the source and the other as the target respectively. Given the source feature $E_s \in \mathbb{R}^{C\times H\times W\times D}$ and the target feature $E_t \in \mathbb{R}^{C\times H\times W\times D}$, BDSA performs $1\times 1\times 1$ convolutions to obtain two triples ($Q_s$, $K_s$, $V_s$) and ($Q_t$, $K_t$, $V_t$), where $K_s$ and $V_s$ are calculated from $E_s$, while $Q_t$, $K_t$, $V_t$, and $Q_s$ are obtained from $E_t$. Secondly, BDSA flattens these features to $(C\times HWD)$, and computes self-attention $A_{self}$ and cross-attention $A_{cross}$:
\begin{equation}
    A_{self} = V_{s}\otimes Softmax(\frac{Q_{s}^T\otimes K_{s}}{\sqrt{C}})^T,\ A_{cross} = V_{t}\otimes Softmax(\frac{Q_{t}^T\otimes K_{t}}{\sqrt{C}})^T,
\end{equation}
where ``$\otimes$'' is the matrix multiply operation. Finally, we concatenate $A_{self}$ and $A_{cross}$, reshape the fused feature to $(C\times H\times W\times D)$, and then use a $3\times3\times3$ convolution to derive the improved target feature $E_{t}^\prime \in \mathbb{R}^{C\times H\times W\times D}$.

\subsection{Target-Aware Modality Weighting (TAMW)}
Fig. \ref{fig: Overview} intuitively shows the TAMW module, which is designed to highlight tumor-related features from multi-modal features to refine segmentation. To be specific, the features $F_{pet}^k, F_{ct}^k \in \mathbb{R}^{C^k\times H^k\times W^k\times D^k}$, generated from MCSA, are concatenated with the up-sampled feature $\Tilde{D}^{k+1} \in \mathbb{R}^{2C^{k+1}\times H^k\times W^k\times D^k}$ of the ${(k+1)}^{th}$ decoder block output to obtain $F^k\in \mathbb{R}^{2(C^{k}+C^{k+1})\times H^k\times W^k\times D^k}$. Next, the up-sampled prediction map $\Tilde{P}^{k+1}$ 
generated from the ${(k+1)}^{th}$ decoder block is used to obtain the foreground feature $F_{fore}^k$ and background feature $F_{back}^k$:
\begin{equation}
    F_{fore}^k = \Tilde{P}^{k+1} \odot F^k,\ 
    F_{back}^k = (1-\Tilde{P}^{k+1}) \odot F^k,
\end{equation}
where $\odot$ denotes the dot product. Then, global average pooling is employed to $F_{fore}^k$ and $F_{back}^k$ to obtain the channel intensity, $I_{fore}^k$ and $I_{back}^k$$\in \mathbb{R}^{2(C^{k}+C^{k+1})\times 1}$. We use an MLP with an activation function $tanh$ to derive the weights for foreground $W_{fore}^k$ and background $W_{back}^k$, obtaining the enhanced feature $F_{enh}^k$:
\begin{equation}
        W_{fore}^k=tanh(MLP(I_{fore}^k)),\ W_{back}^k=tanh(MLP(I_{back}^k)),
\end{equation}
\begin{equation}
        F_{enh}^k=F^k+W_{fore}^k\odot F_{fore}^k+W_{back}^k\odot F_{back}^k\ .
\end{equation}

In TAMW, $W_{fore}^k$ and $W_{back}^k$ select foreground-related and background-related features adaptively for refinement. Take $W_{fore}^k$ as an example, for a specific channel, when the corresponding foreground weight is greater than 0, it is \textit{emphasized} by the network for optimizing tumor-related features; otherwise, it is \textit{weakened}.

\subsection{Loss Function}
H2ASeg outputs five prediction maps corresponding to the five levels of our network. For the prediction map $P^{k}$ of the $k^{th}$ level decoder, we up-sample it to the same size with ground truth $GT$ and calculate binary cross-entropy loss $\mathcal{L}_{bce}$ and Dice loss $\mathcal{L}_{dice}$ \cite{VNet}. The total loss $\mathcal{L}_{total}$ can be formulated as:
\begin{equation}
    \mathcal{L}_{total}=\sum_{k=1}^{k=5} (6-k)(\mathcal{L}_{bce}(P^{k}, GT)+\mathcal{L}_{dice}(P^{k}, GT)).
\end{equation}

\begin{table}[!t]
\centering
\caption{Quantitative comparison (mean$\pm$std) on AutoPET-II and Hecktor2022. The best performance is shown in \textbf{bold}, and the second is \underline{underlined}.}
\label{table: Quantitative}
\resizebox{\textwidth}{!}{
\begin{tabular}{@{}c|c|c|c|c|c|c|c|c@{}}
\hline
\multirow{2}{*}{Methods} & \multicolumn{4}{c}{AutoPET-II} & \multicolumn{4}{|c}{Hecktor2022} \\ 
\cline{2-5}  \cline{6-9}  
& Dice ($\uparrow$) & HD95 ($\downarrow$) & Precision ($\uparrow$) & Recall ($\uparrow$) & Dice ($\uparrow$) & HD95 ($\downarrow$) & Precision ($\uparrow$) & Recall ($\uparrow$) \\ \hline
UNet-3D~\cite{UNet-3D} & $27.50$$\pm16.51$ & $121.35$$\pm49.91$ & $25.51$$\pm20.77$ & $51.42$$\pm 20.38$ & $53.52$$\pm9.66$ & $180.52$$\pm40.81$ & $63.50$$\pm12.07$ & $53.70$$\pm10.95$ \\ 
VNet~\cite{VNet} & $55.07$$\pm16.21$ & $72.90$$\pm43.01$ & $61.94$$\pm18.10$ & \underline{$64.17$}$\pm15.10$ & $51.36$$\pm8.12$ & $185.69$$\pm30.28$ & $54.60$$\pm10.66$ & \underline{$58.50$}$\pm12.13$ \\ 
ResUNet-3D~\cite{ResUNet-3D} & $39.78$$\pm20.53$ & $69.11$$\pm42.95$ & $33.16$$\pm22.53$ & $60.70$$\pm17.15$ & $45.52$$\pm12.83$ & $184.74$$\pm36.19$ & $61.05$$\pm9.01$ & $43.97$$\pm13.52$ \\ 
UNETR~\cite{UNETR} & $39.45$$\pm18.15$ & $92.37$$\pm36.25$ & $42.54$$\pm23.67$ & $52.01$$\pm18.25$ & $46.22$$\pm10.05$ & $207.02$$\pm31.91$ & $57.57$$\pm7.81$ & $47.49$$\pm9.04$ \\ 
SwinUNETR~\cite{SwinUNETR} & $55.75$$\pm17.47$ & $79.18$$\pm31.06$ & $59.73$$\pm20.20$ & $62.37$$\pm16.65$ & $47.29$$\pm7.90$ & $205.99$$\pm29.85$ & $55.77$$\pm7.79$ & $50.03$$\pm7.62$ \\ 
nnUNet~\cite{nnUNet} & $55.93$$\pm12.31$ & $70.12$$\pm28.71$ & $58.37$$\pm14.97$ & $61.20$$\pm15.09$ & $50.04$$\pm6.04$ & \underline{$140.80$}$\pm23.77$ & $56.37$$\pm6.89$ & $57.11$$\pm6.02$ \\ 
NestedFormer~\cite{NestedFormer} & \underline{$57.33$}$\pm16.76$ & $67.38$$\pm34.04$ & \underline{68.85}$\pm21.07$ & $61.48$$\pm18.64$ & $51.65$$\pm7.99$ & $217.33$$\pm30.67$ & $56.85$$\pm9.02$ & $57.17$$\pm8.64$ \\ 
A2FSeg~\cite{A2FSeg} & $52.41$$\pm19.07$ & \underline{$67.24$}$\pm37.75$ & $57.97$$\pm22.03$ & $59.32$$\pm18.65$ & \underline{54.77}$\pm6.56$ & $144.50$$\pm25.84$ & \textbf{70.09}$\pm7.09$ & $51.90$$\pm6.98$ \\ 
SDB~\cite{SDB} & $49.57$$\pm22.06$ & $82.89$$\pm44.24$ & $48.12$$\pm25.78$ & $56.35$$\pm21.50$ & $49.71$$\pm10.21$ & $179.26$$\pm36.49$ & $58.95$$\pm13.88$ & $51.42$$\pm13.10$ \\ 
H2ASeg (Ours) & \textbf{60.03}$\pm14.75$ & \textbf{63.09}$\pm30.44$ & \textbf{68.91}$\pm 16.03$  & \textbf{65.44}$\pm 19.87$ & \textbf{59.69}$\pm 6.14$ & \textbf{131.92}$\pm21.78$ & \underline{68.98}$\pm 7.15$ & \textbf{58.54}$\pm 6.83$\\
\hline
\end{tabular}}
\end{table}

\section{Experiments}
\subsection{Experiments on PET/CT Tumor Segmentation}
\subsubsection{Datasets and Evaluation Metrics.}
We evaluate H2ASeg on two benchmarks: AutoPET-II \cite{AutoPET-II} and Hecktor2022 \cite{Hecktor2022}. \textbf{AutoPET-II} is a whole-body PET/CT tumor image dataset, including 1014 PET/CT scans of 900 patients. Each modality has a $3\times2\times2$ spacing and is already co-registered. Following the recent work \cite{marinov2023mirror}, we first ensure that different scans from one patient do not exist in both the training and testing sets, and then randomly split the dataset in a ratio of training: validation: testing = 6: 2: 2. \textbf{Hecktor2022} is a head and neck tumor image dataset consisting of 524 PET/CT cases collected from 9 centers. Each modality has a $1\times1\times1$ spacing and is registered. Following the recent work \cite{andrearczyk2022segmentation}, we randomly split the data in a ratio of training: validation: testing = 6: 2: 2.

We adopt four widely-used metrics for quantitative evaluation: foreground Dice score (Dice), 95\% Hausdorff distance (HD95) \cite{huttenlocher1993comparing}, Precision, and Recall, where the Dice score, Precision, and Recall are presented as percentages (\%) in experiments. Notably, since there are lesions-free samples in AutoPET-II, we only calculate Dice and HD95 in samples with lesions. To avoid randomness, we ran all experiments four times and reported the averaged metrics.

\subsubsection{Implementation Details.} Our method is implemented in PyTorch 2.0.0 on an NVIDIA GeForce RTX 3090 GPU. We employ the Adam optimizer to optimize the overall parameters with a learning rate of 1e-4 and a weight decay of 1e-5. In experiments, each image is cropped into patches with the size of $128\times 128\times 64$ for training. H2ASeg is trained end-to-end to 200 epochs with a batch size of 2. For MCSA, the window size is set to (8, 8, 4) for each layer of H2ASeg.

\subsubsection{Quantitative and Qualitative Evaluation. }
We compare our H2ASeg with nine advanced methods, including UNet-3D \cite{UNet-3D}, VNet \cite{VNet}, ResUNet-3D \cite{ResUNet-3D}, UNETR \cite{UNETR}, SwinUNETR \cite{SwinUNETR}, nnUNet \cite{nnUNet}, NestedFormer \cite{NestedFormer}, A2FSeg \cite{A2FSeg}, and SDB \cite{SDB}. To ensure fairness, we obtain the experimental results by rerunning their released code based on the same training strategy. 

\begin{figure}[t]
\centering
\includegraphics[width=\textwidth]{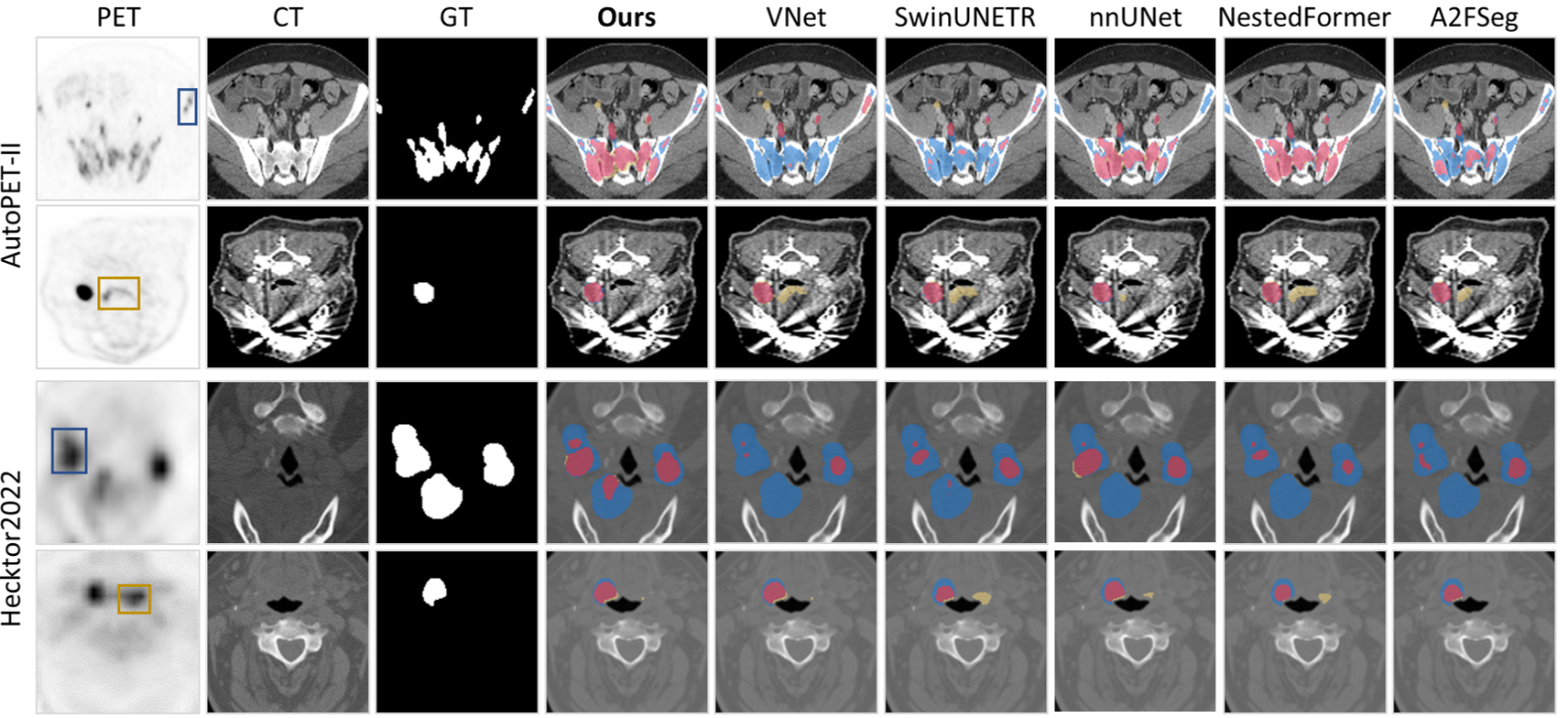}
\caption{Qualitative results of different methods. Boxes in PET highlight areas that are easily misclassified. For segmentation results, the red area is the true positive, the blue is the false negative, and the yellow is the false positive.} 
\label{fig: Qualitative}
\end{figure}

Table \ref{table: Quantitative} lists the quantitative comparison results, in which H2ASeg outperforms other methods on both datasets. On AutoPET-II, due to the limited global modeling ability, CNN-based methods like UNet-3D, ResUNet-3D are easily interfered with by the high metabolic areas, resulting in inferior performance to Transformer-based methods (e.g., UNETR, SwinUNETR). H2ASeg effectively uses Transformer to model the local and global relation across modalities, obtaining state-of-the-art result with Dice of 60.03\%. On Hecktor2022, H2ASeg achieves significant improvements with a gain of $4.98\%$ in Dice to the advanced multi-modal segmentation method A2FSeg. Fig. \ref{fig: Qualitative} provides visible comparison results. It can be seen that H2ASeg shows superior performance in localizing and segmenting tumors of different sizes under various scenes.

\subsection{Ablation Study}
To further evaluate the effectiveness of the proposed modules, we conducted ablation experiments and reported the quantitative results in Table \ref{table: ablation}. From the table, we can find the following. (1) MCSA and TAMW are effective components, which can significantly increase baseline performance by 12.20\% and 10.94\% in Dice respectively. (2) The joint use of MCSA and TAMW can further improve the baseline, achieving the state-of-the-art result on AutoPET-II. This reveals that MCSA and TAMW can synergistically enhance each other, enabling H2ASeg to accurately perform tumor segmentation in PET/CT images.

\begin{figure}[t]
\centering
\resizebox{\textwidth}{!}{
\includegraphics[width=\textwidth]{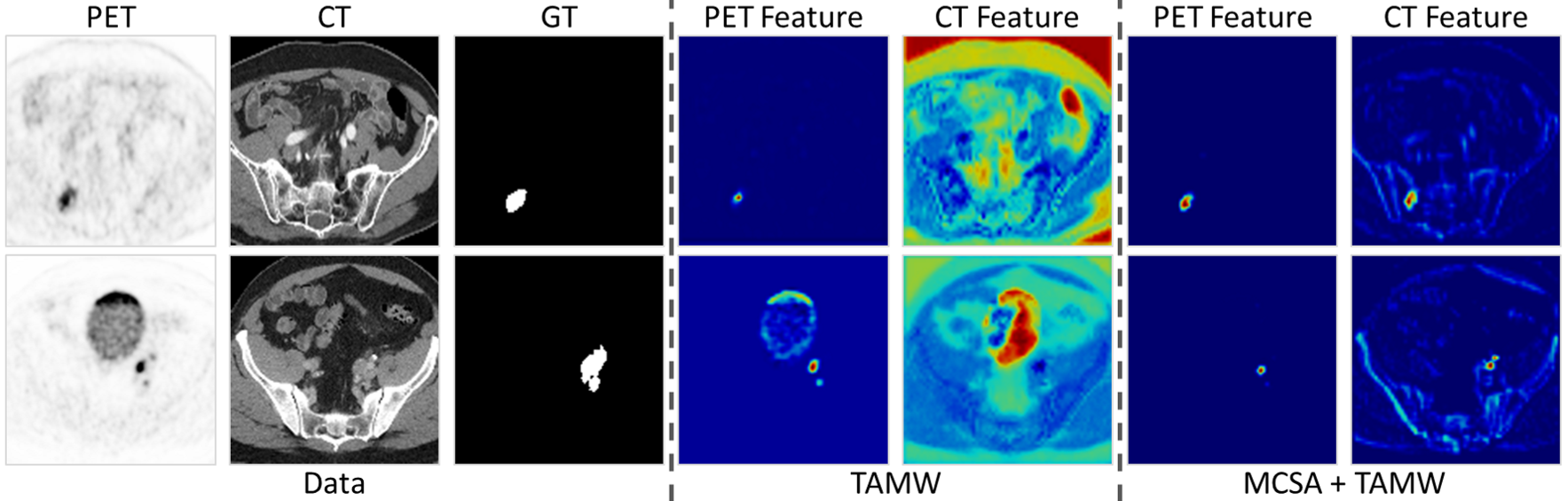}}
\caption{Visualization of feature maps selected by TAMW for foreground emphasis.}
\label{fig: ablation}
\end{figure}

To further prove the effectiveness of TAMW and MCSA modules, we visualize the feature maps (from $F^2_{pet}$ and $F^2_{ct}$) that are selected by TAMW for foreground emphasis in Fig. \ref{fig: ablation}. It can be seen that the independent use of TAMW modules highlights tumor-related features in PET features, but fails to exploit CT features. The introduction of MCSA improves the phenomenon. This is because the modality interaction in MCSA enhances the feature representation of each modality, making the network effectively capture tumor-related features.

Table \ref{table: Emphasis} shows the average weights of PET and CT for foreground emphasis and background emphasis in TAMW at different network depths. For the contribution of PET/CT in foreground emphasis, we observe that the H2ASeg tends to highlight CT features in the shallow layer (texture) and use PET features in deep layers (semantics). These behaviors are consistent with the actual role of PET and CT in clinical diagnosis, which demonstrates the hierarchical interaction of H2ASeg can effectively explore the complementary information contained in PET/CT, utilizing the tumor localization ability of PET and the detailed description ability of CT for precise segmentation.

\begin{figure}[t]
        \centering
        \begin{minipage}[t]{0.4\textwidth}
        \centering
        \tabcaption{Effects of MCSA and TAMW on AutoPet-II.}
        \label{table: ablation}
        \scalebox{0.75}{
        \begin{tabular}{@{}cc|c|c|c|c@{}}
        \hline
        \multicolumn{2}{c|}{Modules} & \multicolumn{4}{c}{AutoPET-II}\\ \hline
        MCSA & TAMW & Dice & HD95 & Precision & Recall \\ 
        \hline
        & & 46.23 & 71.67 & 41.43 & 60.71 \\ 
        \checkmark & & \underline{58.43} & 69.04 & 63.94 & \underline{64.17}  \\
        & \checkmark & 57.19 & \underline{65.91} & \textbf{70.68} & 58.45\\ 
        \checkmark & \checkmark & \textbf{60.03} & \textbf{63.09} & \underline{68.91} & \textbf{65.44} \\
        \hline
        \end{tabular}}
    \end{minipage}
    \hspace{0.1in}
    \begin{minipage}[t]{0.55\textwidth}
        \centering
        \tabcaption{Effects (\%) of PET/CT in TAMW for emphasis at different depths.}
        \label{table: Emphasis}
        \scalebox{0.75}{
        \begin{tabular}{@{}c|c|c|c|c@{}}
        \hline
        \multirow{2}{*}{Depths} & \multicolumn{2}{|c}{Foreground emphasis} & \multicolumn{2}{|c}{Background emphasis}\\ 
        \cline{2-5}
        & PET & CT & PET & CT \\ 
        \hline
        1 & $82.79\pm19.69$ & $95.77\pm9.83$ & $46.44\pm21.61$ & $39.81\pm15.71$ \\ 
        2 & $86.68\pm11.15$ & $65.36\pm34.41$ & $56.26\pm26.83$ & $43.31\pm16.00$\\
        3 & $68.50\pm31.11$ & $53.84\pm21.29$ & $49.61\pm26.89$ & $55.39\pm13.15$\\ 
        4 & $66.83\pm0.29$ & $61.42\pm32.47$ & $51.98\pm20.60$ & $48.15\pm24.95$\\
        \hline
        \end{tabular}}
    \end{minipage}
\end{figure}

\section{Conclusion}
In this paper, we propose a novel architecture H2ASeg that hierarchically models the correlations between modalities, exploiting the complementary information for precise PET/CT tumor segmentation. Specifically, we design a modality-cooperative spatial attention (MCSA) module to adaptively transfer potential information across modalities locally and globally. A target-aware modality weighting (TAMW) module is developed for highlighting tumor-related features for segmentation refinement. Extensive experiments demonstrate the superiority of H2ASeg, with state-of-the-art results on two benchmarks. Moreover, MCSA and TAMW are flexible enough to be embedded into existing architectures with higher efficiency to excavate intrinsic correlations between modalities.

\bibliography{ref.bib}
\end{document}